\documentclass[aps,prb,showpacs,groupedaddress]{revtex4}
\usepackage{amsmath,graphicx}
\usepackage{epsfig}
\usepackage{bm}
\usepackage[next]{inputenc}

\begin{document}
\title{\bf Dynamical and static spin structure factors of Heisenberg antiferromagnet 
on
 honeycomb lattice in the presence of Dzyaloshinskii-Moriya interaction  }
\author{F. Azizi and H. Rezania}
\email[]{rezania.hamed@gmail.com}
\affiliation{Physics Department, Razi Univerity, Kermanshah, Iran}

\begin{abstract}
We have theoretically studied the spin structure factors of Heisenberg model on honeycomb lattice
 in the presence of longitudinal magnetic field, i.e. magnetic field perpendicular to the honeycomb plane, and Dzyaloshinskii-Moriya interaction. 
The possible effects of 
next nearest neighbor exchange constant are investigated in terms of anisotropy in the Heisenberg interactions.
This spatial anisotropy is due to the difference between nearest neighbor exchange coupling constant and next nearest neighbor 
exchange coupling constant.
The original spin model hamiltonian is mapped to a bosonic model via a hard core bosonic transformation where an infinite hard core repulsion
 is imposed to constrain
one boson occupation per site. 
Using Green's function approach, the energy spectrum 
of quasiparticle excitation has been obtained. The spectrum of the bosonic gas has been implemented in order to 
obtain two particle propagator which corresponds to spin structure factor of original Heisenberg chain model Hamiltonian.
The results show the position of peak in the dynamical transverse spin
 structure factor at fixed value for Dzyaloshinskii Moriya interaction moves to higher frequency with magnetic field. 
Also the intensity 
of dynamical transverse spin structure factor is not affected by magnetic field. 
However the Dzyaloshinskii Moriya interaction strength causes to decrease the intensity of dynamical transverse spin structure factor.
The increase of magnetic field does not varied the frequency position of peaks in dynamical longitudinal
spin susceptibility however the intensity reduces with magnetic field.
 Our results show
  static transverse structure factor is found to be monotonically decreasing
 with magnetic field and temperature for different vlaues of next nearest neighbor coupling exchange 
constant.
 \end{abstract}
\date{\today}
\maketitle
\section{Introduction}
Quantum magnetism on geometrically 
two-dimensional (D) frustrated spin systems with $S=1/2$ have lately received massive 
attetions, due to their potential for realizing the quantum spin liquid, a magnetically 
disordered state which respects all the symmetries of the systems, even at absolute 
zero temperature\cite{okuma,zvygain,fihlo,ito,balent}.
The spin model, recently attracted many interests, is the Heisenberg model with first and second
antiferromagnetic exchange interaction in honeycomb lattice.
In sufficiently low dimensional spin systems, the quantum mechanical zero point 
motion can forbid long range magnetic order and produce a quantum 
spin liquid state, a correlated state that breaks no symmetry and possesses topological 
properties, possibly sustaining fractionalized excitations\cite{anderson1,
fazekas,liang,sachdev1,sandvik}. Although the triangular lattice was first theoretically
proposed by Anderson\cite{anderson1} as an ideal benchmark to search for the quantum spin 
liquid, it was soon found that the $S=1/2$
antiferromagnetic Heisenberg model on a triangular lattice is magnetically ordered with a 
120$\textdegree$ arrangement of the spins. Despite the intense activity, only a small number of triangular
materials have been identified as possible candidates for quantum spin liquid behavior
such as th layered organic materials. 
Hence, there is a need to find evidence for quantum spin liquid behavior in more compounds.
Honeycomb lattice materials have attracted lots of 
attetion in recent years
due to their interesting and poorly understood magnetic properties. Inorganic materials
such as Na$_{2}$ Co$_{2}$TeO$_{6}$\cite{lefan},
 BaCo$_{2}$(AsO$_{2}$)$_{2}$ \cite{martin},
 are
examples of honeycomb lattice spin half antiferromagnets. It is important then to understand theoretically
the magnetic properties of interacting localized moments
on the frustrated honeycomb lattice as has been previously done on triangular lattices.
Althogh the numerical evidence for a quantum spin liquid in the half-filled Hubbard model on the 
honeycomb lattice has been questioned\cite{sorella}, exact diagonalization studies on the 
$J_{1}-J_{2}$ Heisenberg model with $S=1/2$ have found evidence for short range 
spin gapped phases for $J_{2}=0.3-0.35$ suggesting the presence of a Resonance Valence Bond
(RVB) state.
Slow neutrons scatter from
solids via magnetic dipole interaction in which the magnetic moment
of the neutron interacts with the spin magnetic moment of electrons
in the solid. The inelastic cross-section for magnetic
neutron scattering from a magnetic system can be expressed in terms
of spin density correlation functions.  
Specially, field
induced effects on the dynamical spin correlation function in low
dimensional quantum spin models have been attracting much interest
from theoretical and experimental point of view in recent
years
\cite{affleck,uimin,dimitriev}.
AF Heisenberg chain in the presence of
axial anisotropy at finite non zero values for magnetic field is a solvable model. Its ground state properties have been investigated within 
Bethe-Ansatz\cite{bethe,faddeev}. Static thermodynamic quantities, eg. the specific heat, the magnetic susceptibility and magnetization have been investigated by 
several methods including thermodynamic Bethe Ansatz, Quantum Monte Carlo and Density Matrix Renormalization Group\cite{klumper,johnston,klumper1}.

Spin-orbit coupling induces both symmetric and
antisymmetric anisotropic properties or Dzyaloshinskii-Moriya spin
anisotropy\cite{moriya} in the exchange coupling between nearest neighbour spins.
A Dzyaloshiskii-Moriya (DM) interaction with a DM vector
perpendicular to the layer produces an easy-plane spin anisotropic
Hamiltonian for honeycomb lattice. DM interaction breaks $SU(2)$
symmetry of the model and reduces it to $U(1)$ symmetry around the
axis perpendicular to the honeycomb plane. This DM interaction is believed to orient the spins in the 2D layer. 
For applied large magnetic fields ($B$) along perpendicular to the plane, the Zeeman term overcomes the antiferromagnetic spin coupling
and the ground state
is a field induced ferromagnetic state with
gapped magnon excitations. Decreasing the magnetic field at the zero temperature,
the magnon gap vanishes at the critical field ($B_{c}$) and a spiral transverse magnetic
ordering develops.
At finite temperature ($T$)  and in the absence of
exchange anisotropy, the frustration leads to an incommensurate
spiral structure for transverse spin component below the critical line in the $B-T$ plane.

Hard core bosonic representation has been introduced to transform a spin
Hamiltonian to bosonic one whereby the excitation spectrum is obtained\cite{auerbach,mahan}.
This mapping between the bosonic gas and original spin model is valid provided to add the hard core
repulsion between particles in the bosonic Hamiltonian\cite{mahan}. An anisotropic exchange
interaction due to DM interaction adds a hopping term for bosonic particles to the main part of bosonic model Hamiltonian. 
Such term leads to different universal behavior in addition to change of critical points
and thermodynamic properties.

In this paper, we intend to find the effects of antisymmetric exchange interaction (DM interaction) and magnetic field on
 dynamical spin structure factor of localized electrons on honeycomb lattice at finite temperature in the field induced gapped spin-polarized phase. 
Specially, we study the effects of Dzyaloshinskii-Moriya interaction strength and next nearest neighbor exchange constant on the
spin structure factors of Heisenberg model on honeycomb structure for
magnetic fields above critical field $B_{c}$.
We have addressed the frequency behaviors of transverse and longitudinal structure factors
 of anisotropic
 antiferromagnetic Heisenberg model on honeycomb lattice for different values of magnetic field above threshold field. 
 We have implemented the hard core boson transformation for spin operators which gives the excitation 
spectrum in terms of many body calculations for bosonic gas\cite{gorkov}. We have used Brueckner approach \cite{gorkov} 
to find the bosonic self-energy
 to get the spinon dispersion relation.
In order to calculate
structure factors of the spin model Hamiltonian including
antisymmetric anisotropic Dzyaloshinskii-Moriya term, we have used one particle excitation spectrum
of hard core bosonic model Hamiltonian.
We obtain both transverse and longitudinal spin structure factors from one and two particle bosonic Green's function. 
Also the effects of next nearest neighbor exchange constant
and DM interaction strength on temperature dependence of transverse 
static spin susceptibility have been investigated in details.
\section{Theoretical formalism}
\begin{figure}
\begin{center}
\epsfxsize=0.8\textwidth
\includegraphics[width=15.cm]{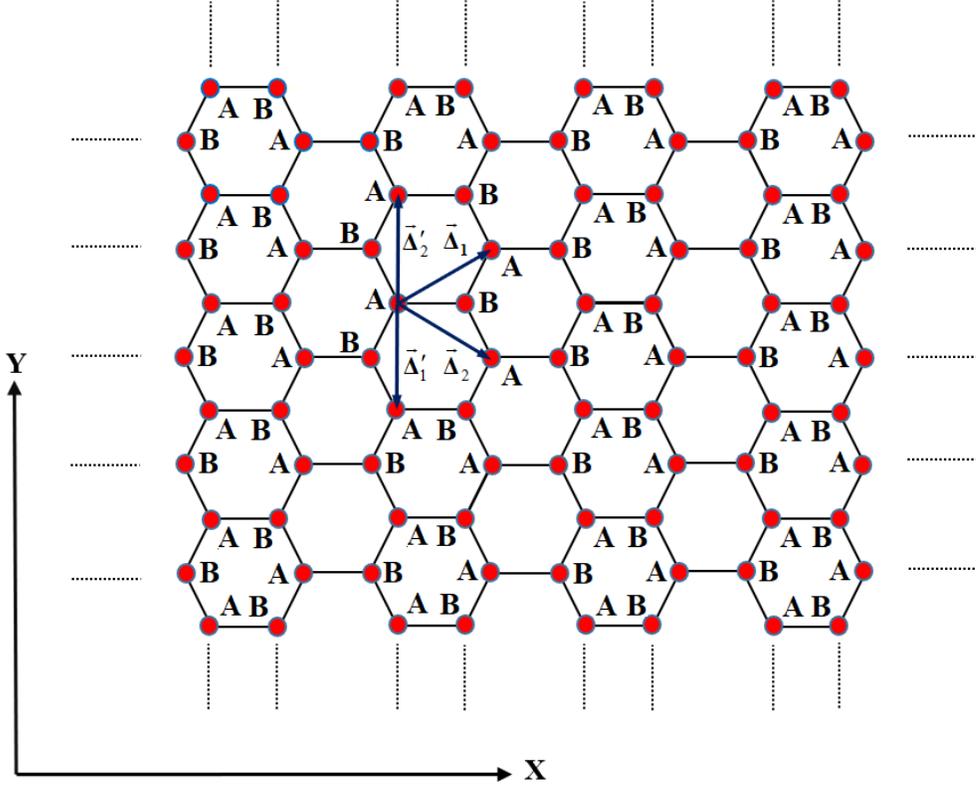}
\small
\begin{flushleft}
\caption{\label{fig111}Crystal structure of honeycomb lattice with two different sublattices.
${\bf a}_{1}$ and ${\bf a}_{2}$ are the primitive unit cell vectors.
The translational vectors $\vec{\Delta}_{1},\vec{\Delta}_{2},\vec{\Delta}'_{1},\vec{\Delta}'_{2}$ 
connecting neighbor unit cells have been shown.}
\end{flushleft}
\end{center}
\end{figure}

The general form of anisotropic spin Hamiltonian on the monolayer honeycomb
lattice which contains antisymmetric exchange interaction
between spins in the presence of a magnetic field is written in the following form
\begin{eqnarray}
H=J\sum_{\langle ij\rangle}{\bf S}_{i}\cdot{\bf S}_{j}+J'\sum_{[ij]}{\bf S}_
{i}\cdot{\bf S}_{j}+{\bf D}.\sum_{\langle ij\rangle}
{\bf S}_{i}\times{\bf S}_{j}-g\mu_{B}B\sum_{i}S_{i}^{z}
\label{e1}
\end{eqnarray}
The lattice structure of honeycomb lattice has been shown in Fig.(\ref{fig111}). 
The symbol $\langle ij\rangle$ and $[ij]$ in Eq.({\ref{e1}}) implies the 
nearest neighbor and next nearest neighbor sites in a honeycomb lattice, respectively.
$J$ is the nearest neighbor coupling constant between spins, while $J'$ denotes the next nearest neighbor
 coupling 
constant between spins. $z$ is thae axis perpendicular to the honeycomb plane.
The antisymmetric spin exchange originating from the relativistic spin-orbit interaction is also known as Dzyaloshinskii-Moriya (DM)
interaction. 
The third term in Eq.(\ref{e1}) with ${\bf D}=(0,0,D)$ describes the 
DM interaction between nearest neighbor sites. 
Also $g\simeq 2.2$ is the gyromagnetic constant and $\mu_B$ denotes the Bohr magneton.
The gyromagnetic constant $g$ value is independent of the type of lattice and gets the same value for honeycomb lattices such as graphene.
This constant is related to the structure of localized electron in the lattice.
 $B$ is the magnetic field strength. The magnetic field is applied as perpendicular to honeycomb plane.
Anisotropy due to DM interaction and Zeeman term violate
 SU(2) symmetry of the isotropic Heisenberg model Hamiltonian.

In order to obtain the bosonic representation of the model hamiltonian
two different bosonic operators are required. 
Therefore spin operators are transformed to bosonic ones as
\begin{eqnarray}
S^{+}_{l,a}=a_{l}\;\;,\;\;S^{-}_{l,a}=a^{\dag}_{l}
\;\;,\;\;S^{z}_{l,a}=1/2-a^{\dag}_{l}a_{l}.\nonumber\\
S^{+}_{l,b}=b_{l}\;\;,\;\;S^{-}_{l,b}=b^{\dag}_{l}
\;\;,\;\;S^{z}_{l,b}=1/2-b^{\dag}_{l}b_{l}.
\label{e2}
\end{eqnarray}
$l$ denotes the unit cell index and $a,b$ label two sublattices. 
$a^{\dag}_{l}(b^{\dag}_{l})$ creates a boson in unit cell with index $l$ on sublattice $a(b)$.
Exploiting above transformation, we have the following one particle bosonic hamiltonian
\begin{eqnarray}
 \mathcal{H}_{bil}&=&\frac{J}{2}\sum_{l,\Delta}(a^{\dag}_{l+\Delta}b_{l}+h.c.)
+\frac{J'}{2}\sum_{l,\Delta'}\Big(a^{\dag}_{l+\Delta'}a_{l}+b^{\dag}_{l+\Delta'}b_{l}\Big)
-\frac{J}{2}\sum_{l}\Big(a^{\dag}_{l+\Delta}a_{l+\Delta}+a^{\dag}_{l}a_{l}+
b^{\dag}_{l+\Delta}b_{l+\Delta}+b^{\dag}_{l}b_{l}\Big)
\nonumber\\&-&\frac{J'}{2}\sum_{l}\Big(a^{\dag}_{l+\Delta'}a_{l+\Delta'}
+a^{\dag}_{l}a_{l}+b^{\dag}_{l+\Delta'}b_{l+\Delta'}
+b^{\dag}_{l}b_{l}\Big)+\frac{D}{2i}\sum_{l,\Delta}\Big(
a^{\dag}_{l+\Delta}b_{l}-b^{\dag}_{l}a_{l+\Delta}\Big)\nonumber\\&+
&g\mu_{B}B\sum_{l}\Big(a^{\dag}_{l}a_{l}+b^{\dag}_{l}b_{l}\Big).
\label{e3}
\end{eqnarray}
Based on Fig.(\ref{fig111}), the symbol $\Delta=\vec{\Delta}_{1},\vec{\Delta}_{2},-\vec{\Delta}_{1},-\vec{\Delta}_{2}$ implies the characters of 
neighbor unit cells including nearest neighbor atomic sites.
The translational vectors  $\vec{\Delta}_{1},\vec{\Delta}_{2}$ connecting
 neighbor unit cells are given by
\begin{eqnarray}
\vec{\Delta}_{1}&=&{\bf i}\frac{\sqrt{3}}{2}+{\bf j}\frac{1}{2}\;\;,\;\;\vec{\Delta}_{2}=
{\bf i}\frac{\sqrt{3}}{2}-{\bf j}\frac{1}{2}
\label{e4}
\end{eqnarray}
Also index $\Delta'=\vec{\Delta}'_{1},\vec{\Delta}'_{2}$ implies the characters of neighbor unit cells including next nearest neighbor atomic sites.
The translational vector $\Delta'_{1},\vec{\Delta}'_{2}$ are given by
\begin{eqnarray}
 \vec{\Delta}'_{1}=-\frac{1}{2}{\bf j}\;\;,\;\;\vec{\Delta}'_{2}=-\vec{\Delta}'_{1}
\label{e4p}
\end{eqnarray}
where the length of until cell vectors is considered to be one.
 In terms of Fourier 
space transformation
of Bosonic operators, the bilinear part of model Hamiltonian is given by
\begin{eqnarray}
 \mathcal{H}_{bil}&=&\sum_{\bf k}\mathcal{A}_{{\bf k}}(a^{\dag}_{{\bf k}}a_{\bf k}+b^{\dag}_{{\bf k}}b_{{\bf k}})+
\sum_{{\bf k}}\Big[\mathcal{B}_{{\bf k}}a^{\dag}_{{\bf k}}b_{{\bf k}}+
\mathcal{B}^{*}_{{\bf k}}b^{\dag}_{{\bf k}}a_{{\bf k}}\Big].
\label{e5}
\end{eqnarray}
The coefficients in the above equation are
\begin{eqnarray}
\mathcal{A}_{{\bf k}}&=&g\mu_{B}B+\phi({\bf k})-\frac{3}{2}J-3J'\;\;,\;\;
\mathcal{B}_{{\bf k}}=\phi'({\bf k})+\phi"({\bf k}),\nonumber\\
\phi({\bf k})&=&J'\Big(cos(\frac{\sqrt{3}}{2}k_{x}+\frac{k_{y}}{2})+cos(
\frac{\sqrt{3}}{2}k_{x}-\frac{k_{y}}{2})+cos(k_{y})\Big),\nonumber\\
\phi'({\bf k})&=&\frac{J}{2}\Big(1+2e^{-i(\frac{\sqrt{3}}{2}k_{x})}cos(\frac{k_{y}}{2})\Big),\nonumber\\
\phi"({\bf k})&=&\frac{D}{2i}\Big(2e^{-i(\frac{\sqrt{3}}{2}k_{x})}cos(k_{y}/2)-1\Big).
\label{e5.1}
\end{eqnarray}
The 
 wave vectors ${\bf k}$ are considered in the first Brillouin zone of the honeycomb lattice.
 Also the quartic part of the model Hamiltonian is obtained as follows
\begin{eqnarray}
 \mathcal{H}_{quartic}=\sum_{k,k',q}(\phi_{{\bf q}}+\phi'_{{\bf q}})a^{\dag}_{
{\bf k}+{\bf q}}b^{\dag}_{{\bf k'}-{\bf q}}b_{{\bf k'}}a_{{\bf k}} \;.
\label{e6}
\end{eqnarray}
In order to reproduce the SU(2) spin algebra, bosonic particles
 must also obey the local hard-core constraint
, i.e only one boson can occupy a single site of lattice.
In terms of Fourier transformation of bosonic operators, 
we can write this part of the Hamiltonian as 
\begin{eqnarray}
 \mathcal{H}_{int}=\mathcal{U}\sum_{k,k',q}(a^{\dag}_{
k+q}a^{\dag}_{k'-q }a_{k'}a_{k}+b^{\dag}_{
k+q}b^{\dag}_{k'-q }b_{k'}b_{k}),
\label{e7}
\end{eqnarray}
where $\mathcal{U}$ implies the strength of the hard core repulsion. In fact $\mathcal{U}$ is the local intrasite integration between
hard core bosons. 
The effect of hard core repulsion part ($\mathcal U \rightarrow \infty$)
of the interacting Hamiltonian in Eq.(\ref{e4}) is dominant
compared with the quartic term in Eq.(\ref{e6}). 
Under taking limit $\mathcal U \rightarrow \infty$ the excitation spectrum of bosonic gas describe the excitation spectrum of original spin model Hamiltonian.

Using a unitary transformation as 
\begin{eqnarray}
 \alpha_{{\bf k}}=u_{{\bf k}}a_{{\bf k}}+v_{{\bf k}}b_{{\bf k}}\;\;,\;\;
\beta_{{\bf k}}=-u_{{\bf k}}a_{{\bf k}}+v_{{\bf k}}b_{{\bf k}},
\end{eqnarray}
the bilinear part of the Hamiltonian is diagonalized as
\begin{eqnarray}
 \mathcal{H}_{bil}&=&\sum_{k}(\omega_{\alpha}({\bf k})\alpha^{\dag}_{{\bf k}}\alpha_{{\bf k}}
+\omega_{\beta}({\bf k})\beta^{\dag}_{{\bf k}}\beta_{{\bf k}})\nonumber\\
\omega_{\alpha}({\bf k})&=&\mathcal{A}_{{\bf k}}+|\mathcal{B}_{{\bf k}}|\;\;,\;\;
\omega_{\beta}({\bf k})=\mathcal{A}_{{\bf k}}-|\mathcal{B}_{{\bf k}}|.
\end{eqnarray}
The Bogoliubov coefficients $u,v$ are given by
\begin{eqnarray}
 u_{\bf k}=\sqrt{\frac{\phi'({\bf k})+\phi"
({\bf k})}{2(\phi'({\bf k})+\phi"
({\bf k}))}}\;\;,\;\;v_{\bf k}=\frac{1}{\sqrt{2}}
\end{eqnarray}
According to bilinear part of
 bosonic model Hamiltonian in Eq.(\ref{e5}),
Fourier transformation of the noninteracting 
 Green's function matrix elements are written in the following form
\begin{eqnarray}
 G^{(0)}_{aa}({\bf k},i\omega_{n})&=&-\int_{0}^{1/(k_{B}T)}d\tau e^{i\omega_{n}\tau}\langle 
\mathcal{T}(a_{{\bf k}}(\tau)a^{\dag}_{{\bf k}}(0))\rangle=
\sum_{j=\alpha,\beta}\frac{1}{2}(\frac{1}{i\omega_{n}
-\omega_{j}({\bf k})})=G^{(0)}_{bb}({\bf k},i\omega_{n}),\nonumber\\
 G^{(0)}_{ab}({\bf k},i\omega_{n})&=&-\int_{0}^{1/(k_{B}T)}d\tau e^{i\omega_{n}\tau}\langle \mathcal{T}(a_{{\bf k}}
(\tau)b^{\dag}_{{\bf k}}(0))\rangle=
u_{{\bf k}}v_{{\bf k}}(\frac{1}{i\omega_{n}-\omega_{\alpha}({\bf k})}-
\frac{1}{i\omega_{n}-\omega_{\beta}({\bf k})}),\nonumber\\
G^{(0)}_{ba}({\bf k},i\omega_{n})&=&-\int_{0}^{1/(k_{B}T)}d\tau e^{i\omega_{n}\tau}\langle 
\mathcal{T}(b_{{\bf k}}(\tau)a^{\dag}_{{\bf k}}(0))\rangle=
u^{*}_{{\bf k}}v_{{\bf k}}(\frac{1}{i\omega_{n}-\omega_{\alpha}({\bf k})}
-\frac{1}{i\omega_{n}-\omega_{\beta}({\bf k})}),
\label{7.777}
\end{eqnarray}
where $\omega_{n}=2n\pi k_{B}T$ denotes the bosonic
Matsubara's frequency. Also $T$ denotes the temperature of the system. 
In Eq.(\ref{7.777}), symbol $\mathcal{T}$ means time-ordering operator.
Since the hard core bosonic repulsion is on-site interaction between bosons,
Wick's theorem\cite{mahan} concludes that only the diagonal elements of bosonic self-energy matrix gets non zero value.
Thus 
we can write down a Dyson's equation for the interacting Green's
function matrix (${\bf G}({\bf k},i\omega_{n})$)  as
\begin{eqnarray}
 {\bf G}({\bf k},i\omega_{n})&=&{\bf G}^{(0)}({\bf k},i\omega_{n})+{\bf G}^{(0)}({\bf k},i\omega_{n}){\bf \Sigma}({\bf k},i\omega_{n}){\bf G}({\bf k},i\omega_{n}),\nonumber\\
{\bf G}^{(0)}({\bf k},i\omega_{n})&=&\left(
                              \begin{array}{cc}
                               G^{(0)}_{aa}({\bf k},i\omega_{n}) &  G^{(0)}_{ab}({\bf k},i\omega_{n}) \\
                              G^{(0)}_{ba}({\bf k},i\omega_{n}) & G^{(0)}_{bb}({\bf k},i\omega_{n}) \\
\end{array}
\right),\nonumber\\
{\bf \Sigma}({\bf k},i\omega_{n})&=&\left(
                              \begin{array}{cc}
                               \Sigma_{aa}({\bf k},i\omega_{n}) & 0 \\
                             0 & \Sigma_{bb}({\bf k},i\omega_{n}) \\
\end{array}
\right),
\label{e7.0}
\end{eqnarray}
where $\Sigma^{Ret}_{aa}({\bf k},\omega)=\Sigma^{Ret}_{bb}({\bf k},\omega)$ is normal  
retarded self-energies diagonal element due to hard core repulsion between bosons.
After using Dyson's series \cite{mahan} for
the interacting Matsubara Green's function matrix in Eq.(\ref{e7.0})
, the low energy limit of single particle retarded Green's function matrix elements are
\begin{eqnarray}
G_{aa}^{sp}({\bf k},\omega)&=& G_{aa}({\bf k},i\omega_{n}\longrightarrow\omega+i0^{+})=
\frac{1}{2}\sum_{j=\alpha,\beta}\frac{Z_{{\bf k}}}{\omega-\Omega_{\alpha}({\bf k})+i0^{+}}=G_{bb}^{sp}({\bf k},\omega)\nonumber\\
G_{ab}^{sp}({\bf k},\omega)&=& G_{ab}({\bf k},i\omega_{n}\longrightarrow\omega+i0^{+})=
u_{{\bf k}}v_{{\bf k}}Z_{{\bf k}}(\frac{1}{\omega-\Omega_{\alpha}({\bf k})+i0^{+}}-
\frac{1}{\omega-\Omega_{\beta}({\bf k})+i0^{+}})\nonumber\\
G_{ba}^{sp}({\bf k},\omega)&=& G_{ba}({\bf k},i\omega_{n}\longrightarrow\omega+i0^{+})=
u^{*}_{{\bf k}}v_{{\bf k}}Z_{{\bf k}}(\frac{1}{\omega-\Omega_{\alpha}({\bf k})+i0^{+}}-
\frac{1}{\omega-\Omega_{\beta}({\bf k})+i0^{+}}).
\label{e12}
\end{eqnarray}
The renormalized excitation spectrum and renormalized single particle weight are given by
\begin{eqnarray}
\Omega_{j=\alpha,\beta}({\bf k})&=&Z_{{\bf k}}(\omega_{j=\alpha,\beta}({\bf k})+\Sigma_{aa}({\bf k},0)),
\nonumber\\
Z_{{\bf k}}^{-1}&=&1-(\frac{\partial Re\Big(\Sigma^{Ret}_{aa}({\bf k},\omega)\Big)}{\partial \omega})
_{\omega=0},
\label{e13}
\end{eqnarray}
 Since the Hamiltonian
$\mathcal{H}_{int}$ in Eq.(\ref{e4}) is short ranged and $\mathcal{U}$ is
large, the Brueckner's approach (ladder diagram summation)
\cite{rezania2008,fetter,gorkov} can be employed for calculating bosonic self-energies in
 the low density limit of
bosonic gas and for low temperature. Firstly, the scattering
amplitude (t-matrix) $\Gamma(p_{1},p_{2};p_{3},p_{4})$ of hard core bosons is
introduced where $p_{i}\equiv({\bf p},ip_{m})_{i}$. ${\bf p}$ is
 momentum vector and $p_{m}\equiv 2m\pi k_{B}T$ implies 
the Matsubara's frequency of the $i$th bosonic
particle.
The basic approximation made in the derivation of $\Gamma(K\equiv
p_{1}+p_{2})$ is that we have considered the diagonal elements of bosonic Green's function for 
making $\Gamma$. It is
 due to the strongly intrasite interaction between bosons. 
According to the Feynman's rules
in momentum space at finite temperature ($T$) and
after taking limit $\mathcal{U}\longrightarrow\infty$, the scattering
amplitude can be written as the following form
\begin{eqnarray}
\Gamma_{aaaa}({\bf K},i\omega_{n})&=&-\Big(k_{B}T\frac{1}
{N}\sum_{{\bf Q},m}
  G^{(0)}_{aa}({\bf Q},iQ_{m}) 
G^{(0)}_{aa}({\bf K}-{\bf Q},i\omega_{n}-iQ_{m})
\Big)^{-1},
 \label{e178}
\end{eqnarray}
where $N$ is the number of unit cells and wave vector ${\bf Q}$ belongs to the
 first Brillouin zone of honeycomb lattice. $k_{B}$ is the Boltzmann constant.
We can perform the summation over Matsubara frequencies $Q_{m}=2m\pi k_{B}T$ 
in the Eq.(\ref{e178}) according to Feynman
rules\cite{mahan} and the final result for scattering amplitude is obtained by
\begin{eqnarray}
\Gamma_{aaaa}({\bf K},i\omega_{n})&=&-\Big(\frac{1}
{N}
\sum_{{\bf Q}} \frac{1}{4}
\sum_{j,j'=\alpha,\beta}(\frac{n_{B}(\omega_{j}({\bf Q}))-n_{B}
(-\omega_{j'}({\bf K}-{\bf Q}))}
{i\omega_{n}-\omega_{j}({\bf Q})-\omega_{j'}({\bf K}-{\bf Q})}
\Big)^{-1},
 \label{e178.5}
\end{eqnarray}
where $n_{B}(x)=\frac{1}{e^{x/k_{B}T}-1}$ describes the Bose-Einstien distribution function. 
The hard core self-energy is
 obtained by using the scattering amplitude
obtained in Eq.(\ref{e178}).
\begin{equation}
\Sigma_{aa}(\textbf{k},i\omega_{n})=-k_{B}T\frac{1}{N}\sum_{p_{m},{\bf p}}
\Gamma_{aaaa}(p,k;p,k)G^{(0)}_{aa}({\bf p},ip_{m})-k_{B}T\frac{1}{N}
\sum_{p_{m},{\bf p}}\Gamma_{aaaa}(p,k;k,p)
G^{(0)}_{aa}({\bf p},ip_{m}). \hspace{3mm}
\label{e190}
\end{equation}
The hard core self-energy is obtained by integration
over internal Matsubara's frequency ($p_{m}$)
\begin{eqnarray}
\Sigma_{aa}({\bf k},i\omega_{n})=\frac{1}{N}\sum_{{\bf p}}\Big(n_{B}(\omega_{\alpha}({\bf p}))
\Gamma({\bf p}+{\bf k},\omega_{\alpha}({\bf p})+i\omega_{n})+
n_{B}(\omega_{\beta}({\bf p}))\Gamma({\bf p}+{\bf k},\omega_{\beta}({\bf p})+i\omega_{n})\Big).
 \label{e211}
\end{eqnarray}
It should be noted that index $m$ is the integer number.  
We can simply obtain the retarded self-energy by analytic continuation 
($i\omega_{n}\longrightarrow\omega+i0^{+}$) of Eq.(\ref{e211}). 
The contribution of $\mathcal{H}_{quartic}$ on the final results is very
small beecause it is composed of quartic terms in the bosonic operators.
It is therefore treated in mean-field approximation. This is equivalent to take only one-loop 
diagrams in to account (first order in $J$).
On the mean field level we have $O_{1}O_{2}=\langle O_{1}\rangle O_{2}+
\langle O_{2}\rangle O_{1}-\langle O_{1}\rangle\langle O_{2}\rangle$ where 
each $O_{1}$ and $ O_{2}$ is a pair of boson operators. We can write for each pair of operators,
\begin{eqnarray}
\langle a^{\dag}_{{\bf k}}a_{{\bf k}'}\rangle&=&-\delta_{{\bf k},{\bf k}'}\frac{k_{B}T}{N}\sum_{n}
G^{(0)}_{aa}({\bf k},i\omega_{n})=\frac{1}{2N}\Big(n_{B}(\omega_{\alpha}({\bf k}))+n_{B}(
\omega_{\beta}({\bf k}))\Big),\nonumber\\
\langle a^{\dag}_{{\bf k}}b_{{\bf k}'}\rangle&=&-\delta_{{\bf k},{\bf k}'}
\frac{k_{B}T}{N}\sum_{n}
G^{(0)}_{ab}({\bf k},i\omega_{n})=\frac{u_{{\bf k}}v_{{\bf k}}}{2N}\Big(n_{B}(\omega_{\alpha}
({\bf k}))+n_{B}(
\omega_{\beta}({\bf k}))\Big).
\label{e211.1}
\end{eqnarray}
Thus, the effect of $H_{quartic}$ is to renormalize the coefficients of bilinear part of 
the Hamiltonian according to the following relations
\begin{eqnarray}
\mathcal{A}_{{\bf k}}&\rightarrow&\mathcal{A}_{{\bf k}}+
(\phi_{{\bf q}=0}+\phi'_{{\bf q}=0})\frac{1}{N}\sum_{{\bf k}}\Big(n_{B}(\omega_{\alpha}({\bf k}))+n_{B}(
\omega_{\beta}({\bf k}))\Big),\nonumber\\
\mathcal{B}_{{\bf k}}&\rightarrow&\mathcal{B}_{{\bf k}}+
(\phi_{{\bf q}=0}+\phi'_{{\bf q}=0})\frac{1}{2N}\sum_{{\bf k}} u_{{\bf k}}
v_{{\bf k}}\Big(n_{B}(\omega_{\alpha}({\bf k}))+n_{B}(
\omega_{\beta}({\bf k}))\Big).
\label{e211.10}
\end{eqnarray}
The renormalized coefficients (Eq.(\ref{e211.10})) will be considered to calculate
the self-energy which are independent of energy (nonretarded in time representation).
\section{Dynamical and static spin structure factors}
The correlation function between spin components of localized electrons at different times
 can be expressed in terms of one and two particle
 bosonic Green's functions. The frequency Fourier transformation of this correlation function produces
 the dynamical spin susceptibility. The frequency position of peaks in dynamical spin susceptibility
 are associated with collective excitation localized electrons described by Heisenberg model Hamiltonian. 
These excitations are related to the spin excitation 
spectrum of localized electrons on Honeycomb structure.
 In the view point of experimental interpretation, the dynamical spin susceptibility of the localized electrons of the system
 is proportional to
 inelastic cross-section for magnetic
neutron scattering from a magnetic system that can be expressed in terms
of spin density correlation functions of the system. In other words the differential
inelastic cross section $d^{2}\sigma/d\Omega d\omega$ is
proportional to imaginary part of spin susceptibilities. $\omega$
denotes the energy loss of neutron beam which is defined as the
difference between incident and scattered neutron energies. The solid angle $\Omega$
implies the orientation of wave vector of scattered neutrons from the localized electrons of the sample.
We can assume the wave vector of incident neutrons is along $z$ direction.
The solid angle $\Omega$ depends on the polar angle between wave vector of scattered neutrons and the wave vector
of the incident neutrons. A note is in order here. The solid angle $\Omega$ is different from the excitation spectrum of hard core bosonic particles
, i.e. $\Omega_{\alpha}({\bf k})$, $\Omega_{\beta}({\bf k})$, in Eq.(\ref{e13}).  
 The frequency position
of peaks in $d^{2}\sigma/d\Omega d\omega$ determines the spin
excitation spectrum of the magnetic system\cite{doniach}.
In order to study the general spin excitation spectrum of the localized electron of the systems, both transverse and longitudial
dynamical spin-spin correlation functions have been calculated.

According to the linear-response theory, the 
components of spin susceptibility tensor are written by
\begin{eqnarray}
\chi_{\eta\gamma}({\bf q},\omega)=i\int_{0}^{+\infty}dt e^{i\omega t}\langle 
[S_{\eta}({\bf q},t),S_{\gamma}(-{\bf q},0)]\rangle&=&lim_{i\omega_{n}
\longrightarrow\omega+i0^{+}}\int_{0}^{1/(k_{B}T)}d\tau e^{i\omega_{n}\tau}\langle 
\mathcal{T}S_{\eta}({\bf q},\tau)S_{\gamma}(-{\bf q},0)\rangle\nonumber\\&=&
\chi_{\eta\gamma}({\bf q},i\omega_{n}\longrightarrow\omega+i0^{+}),
 \label{e952}
\end{eqnarray}
that $\eta,\gamma=x,y,z$ impliy the spatial directions and $\omega_{n}=2n\pi k_{B}T$
 is the bosonic frequency. ${\bf q}$ is the wave vector belonging to the 
first Brillouin one of honeycomb lattice.
 Since each unit cell in honeycomb lattice includes two different
sublattices, the Fourier transformation of spin operators in Eq.(\ref{e952})
is exprssed in terms of both bosonic operators $a_{\bf q}$ and $b_{\bf q}$ as
\begin{eqnarray}
S_{z}({\bf q})=1-\sum_{\bf k}\Big(
a^{\dag}_{{\bf k}+{\bf q}}a_{{\bf k}}+b^{\dag}_{{\bf k}+{\bf q}}b_{{\bf k}}\Big)
\;\;,\;\;S_{x}({\bf q})=\frac{a_{\bf q}+a^{\dag}_{-\bf q}}{2}\;\;,\;\;S_{y}({\bf q})=
\frac{a_{\bf q}-a^{\dag}_{-\bf q}}{2i}.
 \label{65.2}
\end{eqnarray}
Here, 
we define transverse dynamical spin suceptibility, i.e. $\chi_{+-}({\bf q},i\omega_{n})$, in terms of 
components of spin susceptibility tensor, $\chi_{\eta\gamma}({\bf q},i\omega_{n})$,
\begin{eqnarray}
\chi_{+-}({\bf q},i\omega_{n})\equiv
\int_{0}^{1/(k_{B}T)}d\tau e^{i\omega_{n}\tau}\langle 
\mathcal{T}S_{+}({\bf q},\tau)S_{-}(-{\bf q},0)\rangle=\chi_{xx}({\bf q},i\omega_{n})+\chi_{yy}({\bf q},i\omega_{n})+
i\chi_{xy}({\bf q},i\omega_{n})-i\chi_{yx}({\bf q},i\omega_{n}),
 \label{952.2}
\end{eqnarray}
so that spin ladder operators are given as $S_{+}=S_{x}+iS_{y}$
and $S_{-}=S_{x}-iS_{y}$. 
 Using the definitions of Fourier transformation of 
spin operators in Eq.(\ref{65.2}), we can calculate the Matsubara representation of 
transverse dynamical spin suceptibility in terms of one particle bosonic Green's function as
\begin{eqnarray}
\chi_{+-}({\bf q},i\omega_{n})&=&
\int_{0}^{1/(k_{B}T)}d\tau e^{i\omega_{n}\tau}\langle 
\mathcal{T}S_{+}({\bf q},\tau)S_{-}(-{\bf q},0)\rangle
=\int_{0}^{1/(k_{B}T)}d\tau e^{i\omega_{n}\tau}\Big\langle 
\mathcal{T}\Big((a_{\bf q}(\tau)+b_{\bf q}(\tau))(a^{\dag}
_{\bf q}+b^{\dag}_{\bf q})\Big)\Big\rangle\nonumber\\&=&-
\Big(G_{aa}({\bf k},i\omega_{n})+G_{aa}({\bf k},i\omega_{n})
+G_{aa}({\bf k},i\omega_{n})+G_{aa}({\bf k},i\omega_{n})\Big),
\label{e952.2} 
\end{eqnarray}
where the matrix elements of interacting Green's function ($G$) have been expressed in Eq.(\ref{e12}).
Also the Matsubara' form of
 longitudinal dynamical spin susceptibility is written based on two particle Green's function
\begin{eqnarray}
&&\chi_{zz}({\bf q},i\omega_{n})=
\int_{0}^{1/(k_{B}T)}d\tau e^{i\omega_{n}\tau}\langle 
\mathcal{T}S_{z}({\bf q},\tau)S_{z}(-{\bf q},0)\rangle\nonumber\\
&=&\int_{0}^{1/(k_{B}T)}d\tau e^{i\omega_{n}\tau}\Big\langle 
\mathcal{T}\Big(1-\sum_{\bf k}(
a^{\dag}_{{\bf k}+{\bf q}}(\tau)a_{{\bf k}}(\tau)
+b^{\dag}_{{\bf k}+{\bf q}}(\tau)b_{{\bf k}}(\tau)\Big)\Big(1-\sum_{\bf k'}
a^{\dag}_{{\bf k'}-{\bf q}}(0)a_{{\bf k'}}(0)
+b^{\dag}_{{\bf k'}-{\bf q}}(0)b_{{\bf k'}}(0)\Big)\Big\rangle.
\label{e953} 
\end{eqnarray}
After using Wick's theorem\cite{mahan} and performing Matsubara frequency summation rules,
the following expression has been obtained for longitudinal dynamical spin susceptibility
\begin{eqnarray}
\chi_{zz}({\bf q},i\omega_{n})&=&\frac{1}{2}\sum_{\bf k}Z_{{\bf k}+{\bf q}}Z_{\bf k}
\Big[\frac{n_{B}(\Omega_{\alpha}({\bf k}+{\bf q}))-n_{B}(\Omega_{\alpha}({\bf k}))}
{i\omega_{n}+\Omega_{\alpha}({\bf k}+{\bf q}))-\Omega_{\alpha}({\bf k})}
+\frac{n_{B}(\Omega_{\alpha}({\bf k}+{\bf q}))-n_{B}(\Omega_{\beta}({\bf k}))}
{i\omega_{n}+\Omega_{\alpha}({\bf k}+{\bf q}))-\Omega_{\beta}({\bf k})}\nonumber\\
&+&\frac{n_{B}(\Omega_{\beta}({\bf k}+{\bf q}))-n_{B}(\Omega_{\alpha}({\bf k}))}
{i\omega_{n}+\Omega_{\beta}({\bf k}+{\bf q}))-\Omega_{\alpha}({\bf k})}
+\frac{n_{B}(\Omega_{\beta}({\bf k}+{\bf q}))-n_{B}(\Omega_{\beta}({\bf k}))}
{i\omega_{n}+\Omega_{\beta}({\bf k}+{\bf q}))-\Omega_{\beta}({\bf k})}\Big]\nonumber\\
&+&
\frac{1}{4}\sum_{\bf k}Z_{{\bf k}+{\bf q}}Z_{\bf k}
\Big[\frac{n_{B}(\Omega_{\alpha}({\bf k}+{\bf q}))-n_{B}(\Omega_{\alpha}({\bf k}))}
{i\omega_{n}+\Omega_{\alpha}({\bf k}+{\bf q}))-\Omega_{\alpha}({\bf k})}
-\frac{n_{B}(\Omega_{\alpha}({\bf k}+{\bf q}))-n_{B}(\Omega_{\beta}({\bf k}))}
{i\omega_{n}+\Omega_{\alpha}({\bf k}+{\bf q}))-\Omega_{\beta}({\bf k})}\nonumber\\
&-&\frac{n_{B}(\Omega_{\beta}({\bf k}+{\bf q}))-n_{B}(\Omega_{\alpha}({\bf k}))}
{i\omega_{n}+\Omega_{\beta}({\bf k}+{\bf q}))-\Omega_{\alpha}({\bf k})}
+\frac{n_{B}(\Omega_{\beta}({\bf k}+{\bf q}))-n_{B}(\Omega_{\beta}({\bf k}))}
{i\omega_{n}+\Omega_{\beta}({\bf k}+{\bf q}))-\Omega_{\beta}({\bf k})}\Big]\nonumber\\
&\times&
\sqrt{\frac{\phi'({\bf k}+{\bf q})+\phi"({\bf k}+{\bf q})}
{(\phi'({\bf k}+{\bf q})+\phi"({\bf k}+{\bf q}))^{*}}}
\sqrt{\frac{(\phi'({\bf k})+\phi"({\bf k}))^{*}}
{\phi'({\bf k})+\phi"({\bf k})}}.
 \label{e959}
\end{eqnarray}
 The dynamical spin structure factor for both longitudinal and transverse spin
directions are obtained based on retarded presentation of susceptibilities as
\begin{eqnarray}
\chi_{zz}^{Ret}(q,\omega)=\chi_{zz}(q,i\omega_{n}\longrightarrow\omega+i0^+)\;\;,\;\;
\chi_{+-}^{Ret}(q,\omega)=\chi_{+-}(q,i\omega_{n}\longrightarrow\omega+i0^+).
 \label{e962.5}
\end{eqnarray}
so that $\chi_{zz}^{Ret}$ and $\chi_{+-}^{Ret}$ are retarded dynamical spin structure factors
for longitudinal and transverse components of spins, respectively.
It is proportional to the contribution of localized spins 
in the neutron differential cross-section. 
For each ${\bf q}$ the dynamical structure factor has peaks at certain energies which
 represent collective excitations for bosonic  
gas which correspond to the spin excitation spectrum of the original model.

 Static transverse spin
structure factor ($s({\bf q})$) which is a measure of
magnetic long range ordering for spin components along the plane, i.e. transverse direction,
 can be related to imaginary part of
retarded dynamical spin susceptibility using following relation.
\begin{eqnarray}
s_{+-}({\bf q})=
\left\langle S_{+}({\bf q})S_{-}(-{\bf q}) \right\rangle=\chi^{\alpha\alpha}({\bf q},0)&=&
\frac{1}{\beta}\sum_{n}\frac{1}{2\pi}
\int_{-\infty}^{\infty}d\omega\frac{-2Im\chi_{+-}({\bf q},i\omega_{n}\longrightarrow\omega+i0^+)}{i\omega_{n}-\omega}\nonumber\\
&=&\int_{-\infty}^{+\infty}d\omega\frac{n_{B}(\omega)}{\pi}Im\chi_{+-}
({\bf q},i\omega_{n}\longrightarrow\omega+i0^+).
 \label{e963}
\end{eqnarray}
Substituting Eq.(\ref{e12}) into Eq.(\ref{e952.2}) and using Matsubara frequency
summation rules Eq.(\ref{e963}), the final result for transverse static structure factor is given by
  \begin{eqnarray}
   s_{+-}({\bf q})=-Z_{\bf q}\Big[n_{B}(\Omega_{\alpha}({\bf q}))+
n_{B}(\Omega_{\beta}({\bf q}))+\frac{1}{2}\Big(n_{B}(\Omega_{\alpha}({\bf q}))-
n_{B}(\Omega_{\beta}({\bf q}))\Big)\Big(\sqrt{\frac{\phi'({\bf q})+\phi"({\bf q})}
{(\phi'({\bf q})+\phi({\bf q}))^{*}}}+
\sqrt{\frac{(\phi'({\bf q})+\phi"({\bf q}))^{*}}
{\phi'({\bf q})+\phi({\bf q})}}\Big)\Big].
\label{e964}
  \end{eqnarray}
In the next section, the numerical results of dynamical spin structure and static spin 
structure have been presented for various magnetic field and DM interaction strength.
\section{Results and discussions}
The numerical reuslts of spin structure factors 
 of the spin 1/2 Heisenberg model on honeycomb lattice in the field induced spin-polarized phase have been presented in this section. 
The frequency behaviors of dynamical spin susceptibilities and temperature dependence of static spin structure factors are studied.
The effects of both magnetic field and Dzyaloshinskii-Moriya term strength on the behaviors of structure factors are investigated. 
 In the limit of $B/J\longrightarrow\infty$, the ground
state of the original spin model Hamiltonian is a field induced spin-polarized state and a finite energy
gap exists to the lowest excited state.
 The decrease in magnetic field lowers the energy gap
which eventually vanishes at the critical magnetic field ($B_c$).
The effects
of hard core interaction on the excitation spectrum have been obtained by the Green's function approach in the context of
Brueckner's formalism above threshold field $B_{c}$ where the density of bosonic
gas is small. 

The single particle excitation should be found from a self-consistent 
solution of Eqs.(\ref{e13},\ref{e178.5},\ref{e211},\ref{e211.10})
with the substitutions 
$u_{\bf k}\longrightarrow\sqrt{Z_{\bf k}}U_{\bf k}$, $v_{\bf k}\longrightarrow\sqrt{Z_{\bf k}}
V_{\bf k}$, $\omega_{\bf k}\longrightarrow\Omega_{\bf k}$ in the corresponding equations.
The process is started with an initial guess for $Z_{\bf k},\Sigma_{aa}({\bf k},0)$ and
 by using Eq.(\ref{e13}) we find corrected excitation energy.
This is repeated until convergence is reached. 
Using the final values for excitation spectrum
we can calculate the dynamical and static spin structure factors
 by Eqs.(\ref{e952.2},\ref{e959},\ref{e962.5},\ref{e964}). We discuss the numerical results for 
thermodynamic properties in the field induced spin polarized regime where
 energy
spectrum of spin model hamiltonian includes a finite energy gap
between ground state and first excited state. Therefore as long as
excitation spectrum $\Omega_{{\bf Q}_{0}=(0,4\pi/3)}$ has non zero values, the system
preserves its gapped spin polarized phase. In the following numerical results for both static and dynamical spin susceptibilities
the wave vector has been fixed at generic wave vector ${\bf q}_{0}=(2\pi/\sqrt{3},0)$.

\begin{figure}
\includegraphics[width=10cm]{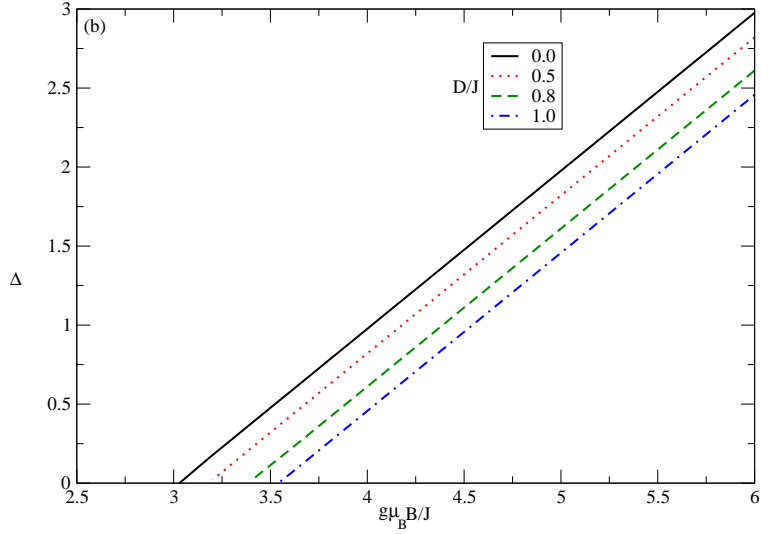}
\caption{\label{fig222}Energy gap ($\Delta$) versus
magnetic field ($g\mu_{B}B/J$) for $J'/J=0.2$ and different values
$D/J$ by setting $kT/J=0.05$.
The change in the critical magnetic field (where the gap vanishes)
for various anisotropies is remarkable.}
\end{figure}

The effect of DM interaction strength, $D$, on critical magnetic field has been studied
in Fig.(\ref{fig222}). In this figure,
energy gap ($\Delta$) versus magnetic field $g\mu_{B}B/J$
for different values of Dzyaloshinskii-Moriya interaction strength
 $D/J$ for $J'/J=0.2$ by setting $k_{B}T/J=0.05$ has been plotted.
Fig.(\ref{fig222}) shows that the energy gap
vanishes as the magnetic field approaches the critical value for
$g\mu_{B}B_{c}/J$. For all values of $D/J$, the gap vanishes at the critical point
 $g\mu_{B}B_{c}/J$
where the transition from gapped spin liquid phase to the gapless
one occurs. Moreover the critical field tends to higher value with
increase of $D/J$ according to Fig.(\ref{fig222}).

\begin{figure}
\includegraphics[width=10cm]{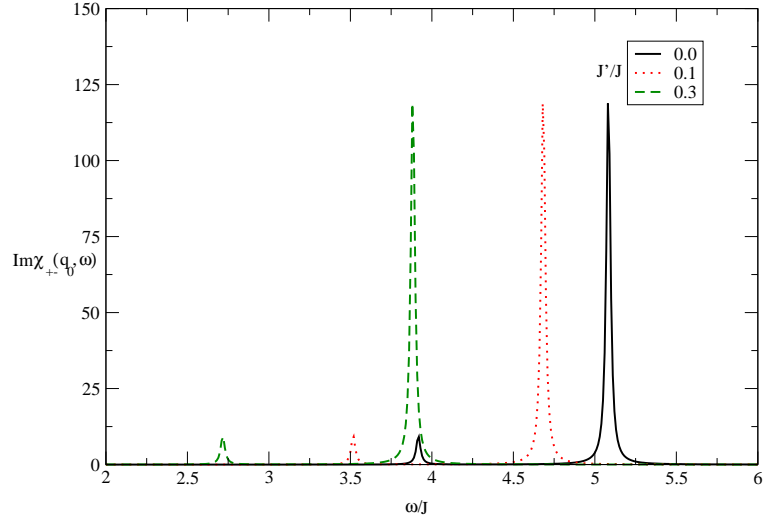}
\caption{\label{fig1} Transverse dynamical spin structure factor
 $Im \chi_{+-}({\bf q}_{0},\omega)$ versus normalized frequency $\omega/J$ 
for different values of next nearest neighbor coupling exchange constant $J'/J$ for $g\mu_{B}B/J=6.0$ and $D/J=0.2$.
 Also normalized temperature is set to $kT/J=0.03$.}
\end{figure}

In Fig.(\ref{fig1}), the dynamical transverse spin structure factor, i.e. 
$Im \chi_{+-}({\bf q}_{0},\omega)$, has been plotted
as a function of frequency $\omega/J$ for different next nearest neighbor coupling 
exchange constants, namely
 $J'/J=0.0,0.1,0.3$, at fixed magnetic field $g\mu_{B}B/J=6.0$ for $D/J=0.2$.
For each value for $J'/J$, there are two sharp peaks in imaginary part of transverse
 dynamical spin susceptibility at low and high frequencies. 
These frequency positions of sharp peaks describe the transverse
 magnetic excitation spectrum of the localized electrons on honeycomb structure.
As shown in Fig.(\ref{fig1}), 
the lower frequency peak have lower height in comparison with the higher frequency peak for 
each value of $J'/J$. The height of each peak corresponds to the intensity of scattered neutrons
from the sample which describes the inelastic scattering cross section of neutron beam. 
Moreover the frequency of each peak is equal to the difference between scattered
neutron beam energy and incident neutron energy beam. This energy difference is absorbed
 by the localized
electrons which causes to magnetic excitation of transvesre components of spins.
According to Fig.(\ref{fig1}), it is clearly observed that frequency position of peaks in the 
imaginary part of transverse spin susceptibility, $Im \chi_{+-}({\bf q}_{0},\omega)$, goes to 
lower values with increase of next nearest
neighbor exchange constant $J'/J$. It can be understood from this fact that the energy gap 
between ground state and first excited state decreases with $J'$
and consequently the magnetic excitation appears at lower frequency.
Moreover the Fig.(\ref{fig1}) implies the intensity of scattered neutron beam from the 
sample is independent of $J'/J$ and the heights of peaks in transverse spin structure factor
 are the same.
The emergence of peaks in the dynamical spin susceptibilities
 of Heisenberg chain have been shown for magnetic fields below threshold one 
using quantum Monte Carlo method based on the stochastic series expansion and maximum entropy method\cite{brenig}. 
At threshold magnetic field, the energy gap closes so that long range magnetic ordering 
develops for magnetic field below threshold one.
Based on Fig.(\ref{fig222}), the normalized threshold magnetic field is about $g\mu_{B}B_{c}/J\approx3$.
The estiamtion value for exchange coupling constant is around $J\approx1 meV$. Thus one can obtain the threshold magnetic field $B_{c}\approx 60$ Tesla.
 However the magnetic field for
 theoretical studies of Heisenberg model Hamiltonian on lattice structures has been considered as normalized value $g\mu_{B}B/J$\cite{langer,zotos}.
In theoretical studies the qualitative behavior of physical parameters in terms of magnetic field is more important.
Also the normalized magnetic field value reported in the theoretical works is in the region $0<g\mu_{B}B/J<5.0$.
Also dynamical structure factor of two dimensional antiferromagnet in high fields has been investigated within spin-wave theory 
using self-consistent approach beyond
the $1/S$ approximation\cite{fuhrman}. The emergence of sharp peak in the frequency behavior of spin structure factor has been predicted in this work in analogy to 
our work.  

\begin{figure}
\includegraphics[width=10cm]{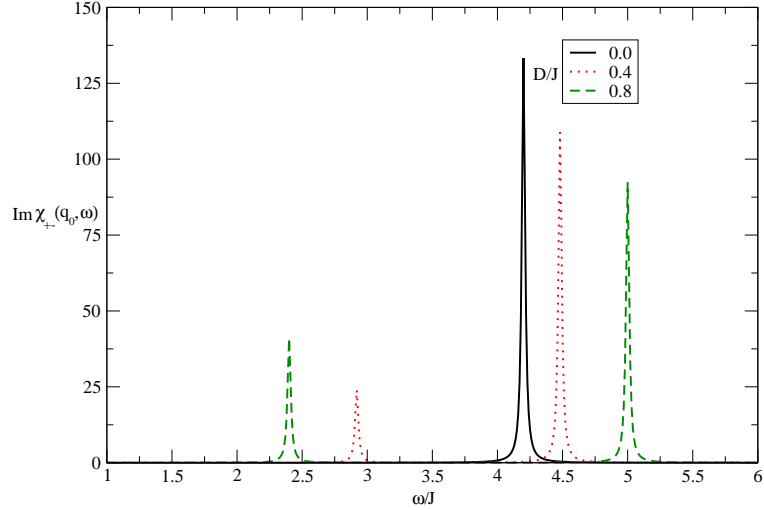}
\caption{\label{fig2} Transverse dynamical spin structure factor $Im \chi_{+-}({\bf q}_{0},\omega)$ versus normalized frequency $\omega/J$ 
for different values of Dzyaloshinskii-Moriya interaction strength $D/J$ for $g\mu_{B}B/J=6.0$ and $J'/J=0.2$.
 Also normalized temperature is set to $kT/J=0.03$.}
\end{figure}

The effect of Dzyaloshinskii-Moriya interaction strength, $D$, as anisotropy parameter on the 
frequency behavior of imaginary part of dynamical transverse spin susceptibility
has been shown in Fig.(\ref{fig2}) for $J'/J=0.2$ at fixed temperature $k_{B}T/J=0.03$.
Based on experimental works on inorganic Heisenberg magnetic materials\cite{lefan,martin} the real value for $k_{B}T/J=0.03$ is expected.
This normalized temperature $k_{B}T/J=0.03$ corresponds to $T\approx0.2 K$.
Also longitudinal magnetic field has been fixed at $g\mu_{B}B/J=6.0$.
This figure includes the sharp peaks at finite frequencies for each value of $D$.
In the absence of $D$ there is no the low frequency excitation mode and we see a collective
magnetic excitation mode at $\omega/J\approx4.2$. 
In the presence of finite $D/J$, an excitation mode appears at low frequencies so that 
the position of peaks tends to lower values with increase of $D/J$. This is a signature of the decrease of energy gap with increase of Dzyaloshinskii
Moriya interaction strength.
Moreover it is cearly observed the high frequency of excitation mode goes to higher values 
with Dzyaloshinskii-Moriya interaction strength. Another novel feature in Fig.(\ref{fig2}) is 
the reduction of intensity of scattered neutrons at frequencies above 4.0 with increase of 
$D/J$. 
 
We have also studied the effect of longitudinal magnetic
field on
 frequency behavior of imaginary part of dynamical transverse spin susceptibility.
In Fig.(\ref{fig3}), $Im \chi_{+-}({\bf q}_{0},\omega)$ has been plotted versus normalized
frequency $\omega/J$ for different values of magnetic field $g\mu_{B}B/J$ above critical field
 for $D/J=0.2$
for $J'/J=0.2$ at fixed temperature $k_{B}T/J=0.03$.
As shown in Fig.(\ref{fig3}), the frequency position of sharp
peaks that describe the magnetic excitation mode of the localized electrons moves to higher 
values upon increasing magnetic field. It can be understood from this fact that energy gap between 
ground state and first excited state in bosonic gas spectrum enhances with $g\mu_{B}B/J$.
Also the height of peaks that is proportional to the intensity of scattered neutrons from the 
electrons of the sample is independent of magnetic field.

In Fig.(\ref{fig4}), the longitudinal dynamic structure factor $Im \chi_{zz}({\bf q}_{0},\omega)$has been shown
as a function of frequency $\omega/J$ for different amounts of 
$J'/J$ for $D/J=0.2$ and $g\mu_{B}B/J=6.0$. This figure shows that there is not sharp peaks 
in imaginary part of longitudinal dynamical spin susceptibility in contrast of 
transverse susceptibility case. In fact there is no 
 well defined magnetic collective excitation spectrum for transverse
components of spins. $Im \chi_{zz}({\bf q}_{0},\omega)$ has two peak at low and high frequencies
for each value of $J'/J$. High frequency peaks appear at frequency $\omega/J=2.0$ however
the intensity of these peaks increase with $J'/J$ according to Fig.(\ref{fig4}).
Low frequency peaks moves to lower values upon increase of next nearest neighbor
exchange constant. Moreover the intensity of these peaks enhances with $J'/J$ similar to the case
of high frequency peaks. Also it is clearly observed that $Im \chi_{zz}({\bf q}_{0},\omega)$
disappears in frequency region above 3.5 for all values of $J'/J$. 
There are the clear differences between numerical results in Figs.(\ref{fig3},\ref{fig4}). 
Fig.(\ref{fig3}) shows very sharp peaks for transverse spin susceptibility
 while the broaden peaks is clearly observed for 
longitudinal dynamical spin susceptibility in Fig.(\ref{fig4}). Also the magnitude of peaks in transverse and longitudinal susceptibilities
are clearly different. These differences arises from this point that magnetic excitation spectrum
for different spin components of localized electrons are different.
In the presence of magnetic field along $z$ direction, 
the magnetic excitation for longitudinal components of spins
is not well performed. This fact can be understood form this point 
the applying magnetic field causes to allignment of spin direction along $z$ direction.
In other words the applied magnetic field suppresses to excite the spin components along 
$z$ direction.
Thus there is no well defined excitation spectrum for $z$ components of localized spins.
In other words the dynamical spin susceptibility $Im \chi_{zz}({\bf q}_{0},\omega)$
includes only broaden peaks according to Fig.(\ref{fig4}). This is not the case for magnetic excitation spectrum
of transverse components od localized spins. These components are
 perpendicular to the magnetic field direction so that the excitation of transverse components
of spins is well performed. This point demonstrates the frequency behavior of transverse
dynamical susceptibility $Im \chi_{+-}({\bf q}_{0},\omega)$ has sharp well defined peaks.
The difference between the magnitudes of peaks for transverse and longitudinal 
spin susceptibilities
is related to intensity of scattered neutrons from the sample.
The applying the magnetic field leads to different intensity of scattered neutron beam for 
transverse and longitudinal components of localized spins.

The effect of Dzyaloshinskii-Moriya interaction strength, $D$, on frequency behavior
of $Im \chi_{zz}({\bf q}_{0},\omega)$ for magnetic field $g\mu_{B}B/J=6.0$ at $J'/J=0.2$ has been 
shown in Fig.(\ref{fig5}). For each value of $D/J$, there is two separate peaks in 
 the imaginary part of longitudinal
dynamical spin susceptibility versus frequency. The first peak appears at frequency 
$\omega/J\approx1.0$ for all values of $D/J$ while the position of the second peak
 moves to higher frequency with increase of $D/J$ based on Fig.(\ref{fig5}).
Another feature in this figure is the enahncement of the amount of 
$Im \chi_{zz}({\bf q}_{0},\omega)$ with $D/J$ at fixed value of frequency.

In Fig.(\ref{fig6}), we present the imaginary part of
 longitudinal dynamical spin susceptibility of 
localized electrons on honeycomb lattice versus normalized frequency $\omega/J$
for different magnetic fields, namely $g\mu_{B}B/J=6.0,6.1,6.2,6.3$
 for fixed values of $J'/J=0.2$ and $D/J=0.2$.
 All plots in Fig.(\ref{fig6}) show the position of peaks is 
almost the same for various values of magnetic field
 however the intensity of peaks reduces with magnetic field strength. This reduction of 
intensity can be justified based on the increase of energy gap with $g\mu_{B}B/J$.

The behavior of transverse static spin structure
factor $s_{+-}({\bf q}_{0})$ versus normalized magnetic field 
$g\mu_{B}B/J$ for different values of $J'/J$ 
has been presented in Fig.(\ref{fig7}). Temperature and DM interaction strength have been fixed 
at $k_{B}T/J=0.03$ and $D/J=0.2$, respectively. 
This figure implies static spin structure for spin components parallel to the honeycomb plane shows a monotonic 
deceasing behavior in terms of magnetic field 
for all values of $J'/J$. This behavior can be understood from this fact that 
magnetic field causes to reduce the magnetic ordering for transverse components of 
spins which decreases transverse static structure factor. 
In addition, at fixed vales of magnetic field, lower next nearest neighbor exchange constant
causes lower energy gap and consequently higher values in transverse structure factor. 
    
The effect of Dzyaloshinskii-Moriya interaction strength on magnetic
field dependence of transverse static spin structure
factor $s_{+-}({\bf q}_{0})$ has been shown in Fig.(\ref{fig8}). 
Here transverse structure factor decreases with magnetic field as shown 
in Fig.(\ref{fig8}). A monotonic decreasing behavior for transverse static structure factor 
has been observed for each DM interaction strength. Moreover transverse structure
factor enhances with increasing $D/J$.

We have also studied the dependence of transverse static spin structure factor on normalized
temperature.
In Fig.(\ref{fig9}), we have plotted the static structure factor of localized electrons on honeycomb
lattice as a function of normalized temperature for different values of $J'$ for $D/J=0.2$.
The magnetic field is set to $g\mu_{B}B/J=6.0$.
However the real value of this magnetic field is higher than 60 Tesla, the qualitative behavior of static structure factor in the 
presence of finite magnetic field is the main aim of this study. Thus a finite non zero normalized magnetic field, for example $g\mu_{B}B/J=6.0$,
has been considered so that the spin liquid phase is valid in this situation.
 According to this figure, structure factor is less
temperature dependent for $k_{B}T/J<0.75$ because the thermal fluctuations has no effect on
magnetic ordering. Upon more increase of temperature the transverse static structure factor
decreases for each value of $J'/J$. Another novel feature in Fig.(\ref{fig9}) is the reduction
of structure factor with decrease of $J'/J$ at fixed temperature. 
This arises from this fact that energy gap increases with decreasing $J'$.

Finally, static transverse spin structure factor as a function of temperature for various
DM interaction parameter has been studied in Fig.(\ref{fig10}) for $g\mu_{B}B/J=6.0$.
Transverse static structure factor has a constant value for temperatures below 1.0. Upon increasing 
temperature above 1.0, thermal fluctuations becomes considerable and consequently the structure
factor decreases. For $k_{B}T/J>3.0$ structure factor is approximately independent of DM interaction
strength where all plots fall on each other in this temperature region. Moreover
at fixed temperature below 3.0
the static structure factor rises with DM interaction strength due to decrease of energy
gap with $D/J$.

We would like to add few comments on the possibility of quantum spin liquid on a Heisenberg 
honeycomb antiferromagnet. Quantum spin liquid phase originates from the finite band gap in excitation spectrum
of Heisenberg model hamiltonian on the lattice. Such finite band gap leads to the exponential behavir
for spatial correlation function between localized spins on the lattice.
 The investigation of quantum spin liquid phase on honeycomb lattice in the absence of 
external magnetic field has been performed 
using theoretical and numerical methods\cite{sorella,assaad,wang,clark,cabra}.
However there is finite
 energy gap in excitation spectrum of Heisenberg model on honeycomb lattice for magnetic
field above critical one. As in Fig.(\ref{fig222})
shown, the energy gap gets non zero value fpr magnetic fields higher than critical one.
Based on Fig.(\ref{fig222}), our results show the threshold magnetic field, where 
energy gap closes, is in the $g\mu_{B}B/J>3.0$. This energy gap in excitation spectrum arises from
Zeeman splitting effect duen spin degrees of freedom. This gapped spin liquid
 phase due to applying magnetic
field is named field induced ferromagnetic phase.
At any temperature, a non zero threshold
 magnetic field is found so that the energy gap in the spectrum 
appears and quantum spin liquid phase develops for magnetic fields above threshold one. 
The inelastic scattering cross section of neutrons from the localized spins for each 
magnetic phase is applied to 
study the collective magnetic excitation modes of the system. In fact dynamical spin susceptibility
corresponding to the inelastic scattering cross section has no infromation regarding the magnetic
phase of the system. Of course the divergence behavior of static spin
 structure factor can be an evidence for phase transition in the system.

\section{Conclusions}
In conclusion, we have the effects of both nexnearest neighbor coupling exchange constant and spin-orbit coupling on the 
magnetic structure factors and collective magnetic modes of Heisenberg model Hamiltonian on honeycomb lattice.
Using hard core bosonic transformation, the excitation spectrum of the model hamiltonian has been found.
The dynamical transverse and longitudinal spin susceptibilities have been calculated using Wick's theorem and 
excitation spectrum of the model. The results show the transverse spin susceptibility includes the sharp well defined
peaks while the longitudinal spin susceptibility has sharp peaks in its frequency dependence. 
Also the behavior of static spin susceptibilities in terms of magnetic field and temperature
has been investigated for different spin-orbit coupling strengths.
 Based on our results, the transverse static spin structure factor decreases with increae 
of magnetic field for different next nearest neighbor exchange constants and spin-orbit 
coupling interactions.


\begin{figure}
\includegraphics[width=10cm]{fig3.eps}
\caption{\label{fig3}Transverse dynamical spin structure factor $Im \chi_{+-}({\bf q}_{0},\omega)$ versus normalized frequency $\omega/J$ 
for different values of longitudinal magnetic field $g\mu_{B}B/J$ for $D/J=0.2$ and $J'/J=0.2$.
 Also normalized temperature is set to $kT/J=0.03$.}
\end{figure}
\begin{figure}
\includegraphics[width=10cm]{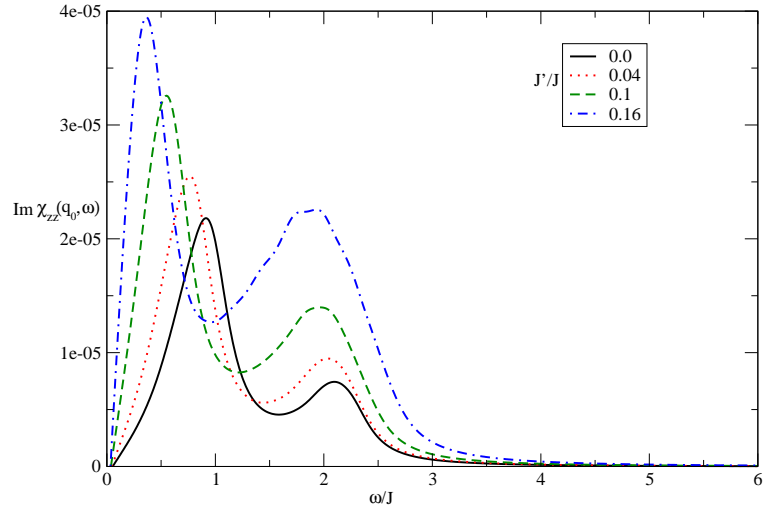}
\caption{\label{fig4}Longitudinal dynamical spin structure factor $Im \chi_{zz}({\bf q}_{0},\omega)$ versus normalized frequency $\omega/J$ 
for different values of next nearest neighbor coupling exchange constant $J'/J$ for $g\mu_{B}B/J=6.0$ and $D/J=0.2$.
 Also normalized temperature is set to $kT/J=0.03$.}
\end{figure}
\begin{figure}
\includegraphics[width=10cm]{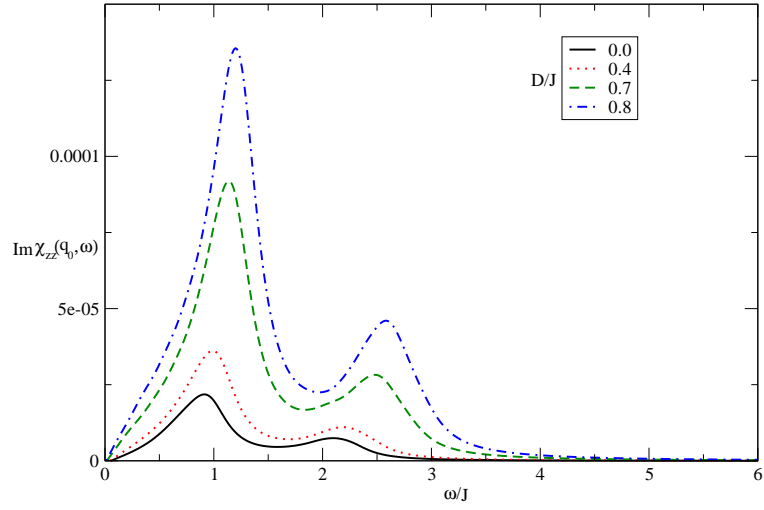}
\caption{\label{fig5}Longitudinal dynamical spin structure factor $Im \chi_{zz}({\bf q}_{0},\omega)$ versus normalized frequency $\omega/J$ 
for different values of Dzyaloshinskii-Moriya interaction strength $D/J$ for $g\mu_{B}B/J=6.0$ and $J'/J=0.2$.
 Also normalized temperature is set to $kT/J=0.03$.}
\end{figure}
\begin{figure}
\includegraphics[width=10cm]{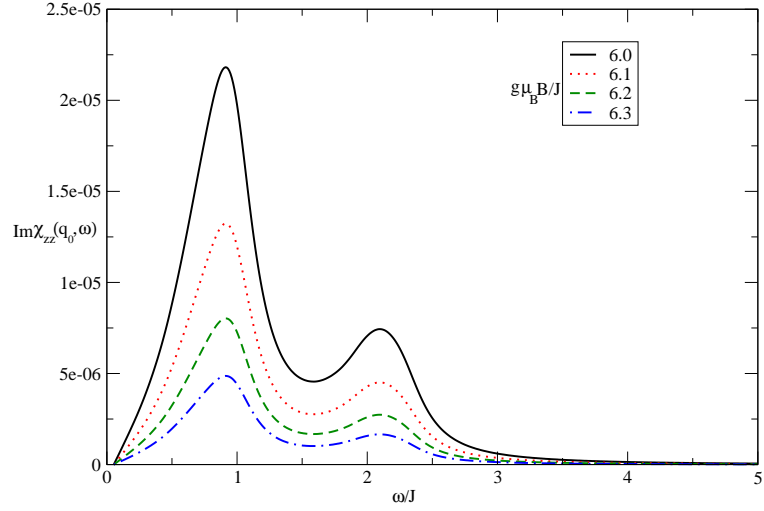}
\caption{\label{fig6}Longitudinal dynamical spin structure factor $Im \chi_{zz}({\bf q}_{0},\omega)$ versus normalized frequency $\omega/J$ 
for different values of longitudinal magnetic field $g\mu_{B}B/J$ for $D/J=0.2$ and $J'/J=0.2$.
 Also normalized temperature is set to $kT/J=0.03$.}
\end{figure}
\begin{figure}
\includegraphics[width=10cm]{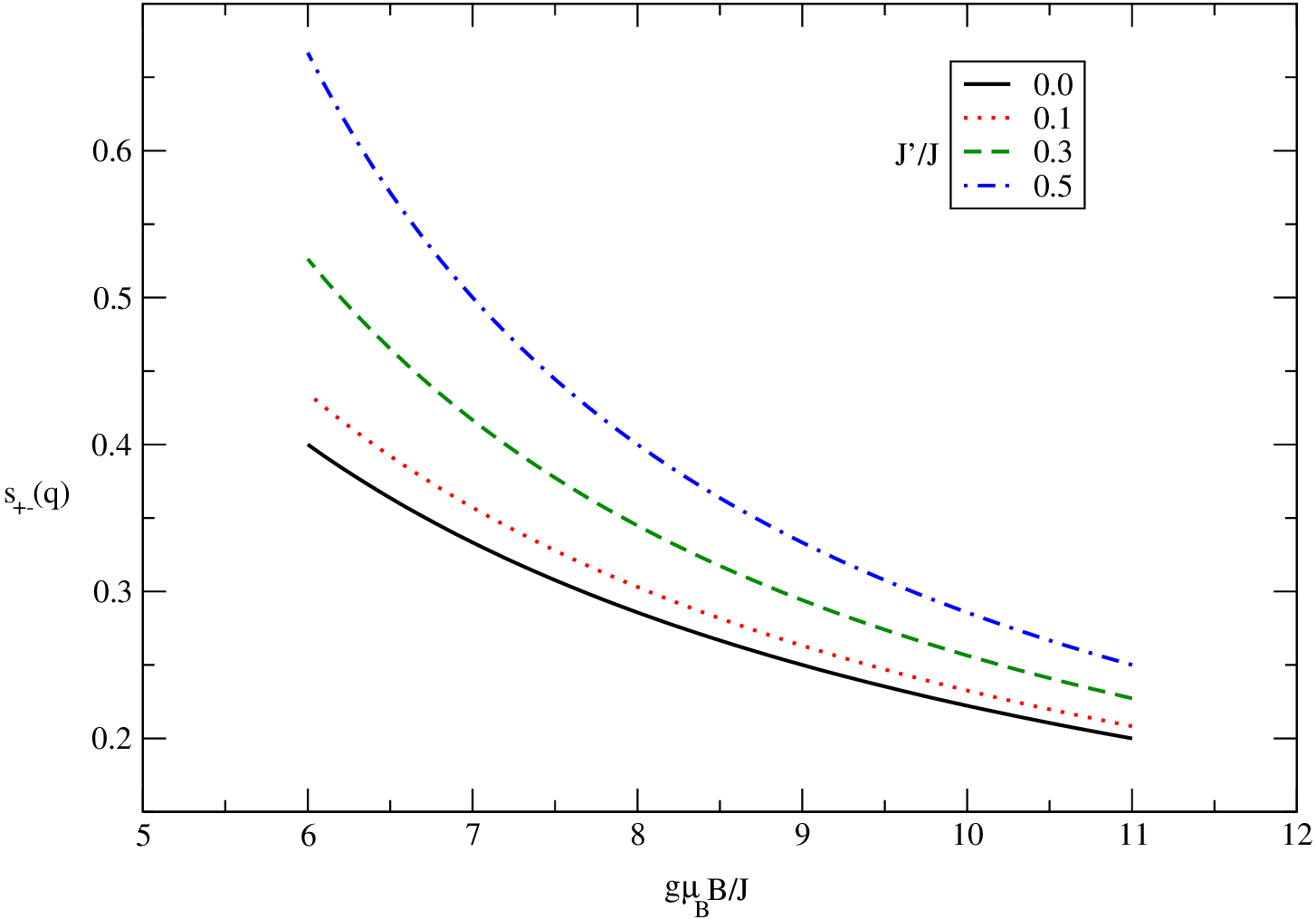}
\caption{\label{fig7}Transverse static spin structure factor $s_{+-}({\bf q}_{0},B)$ as a function normalized magnetic field $g\mu_{B}B/J$ 
for different values of next nearest neighbor exchange coupling constant $J'/J$ for $D/J=0.2$. Also normalized temperature is set to $kT/J=0.03$.}
\end{figure}
\begin{figure}
\includegraphics[width=10cm]{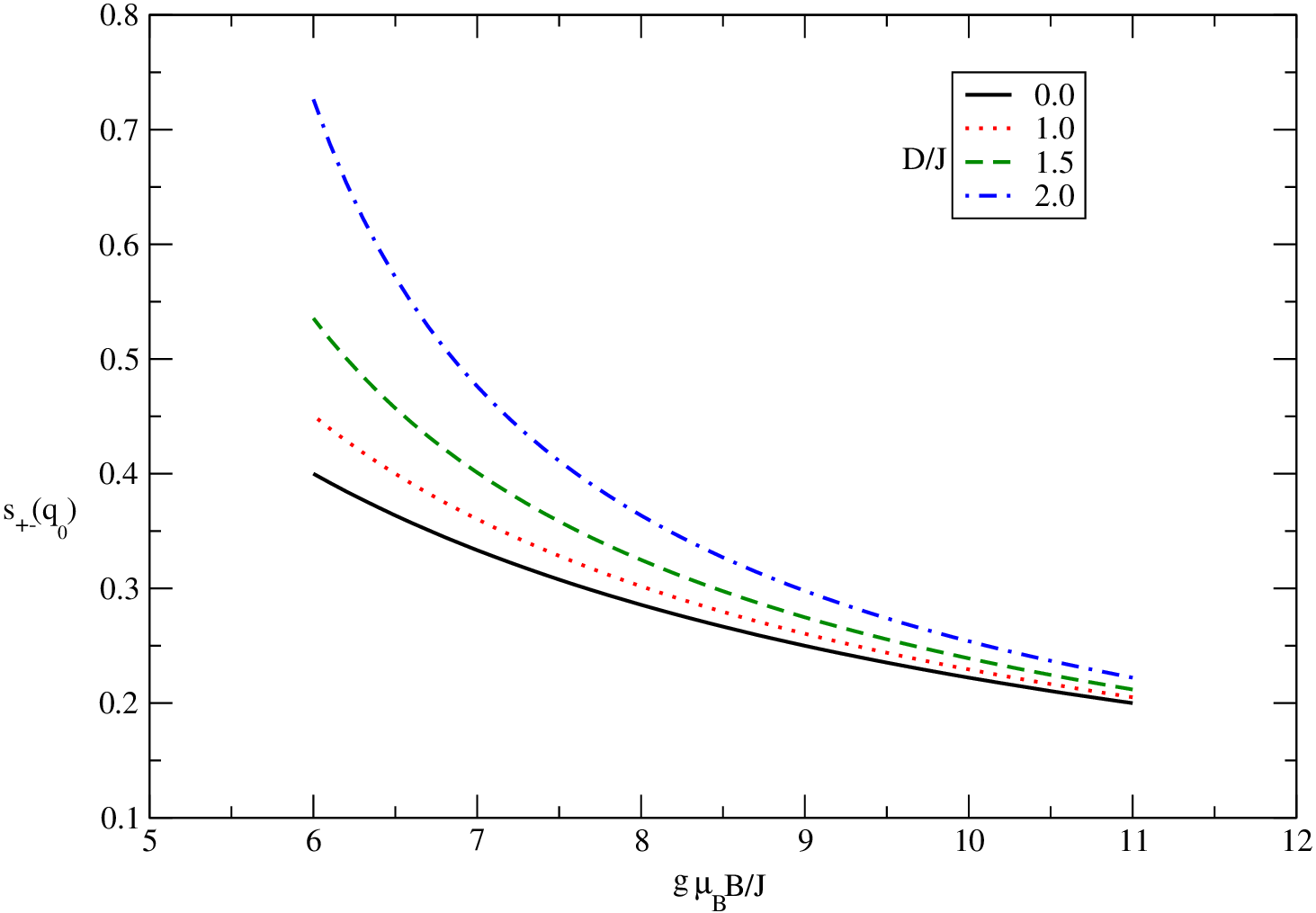}
\caption{\label{fig8}Transverse static spin structure factor $s_{+-}({\bf q}_{0},B)$ as a function normalized magnetic field $g\mu_{B}B/J$ 
for different values of Dzyaloshinskii-Moriya interaction strength $D/J$ for $J'/J=0.2$. Also normalized temperature is set to $k_{B}T/J=0.03$.}
\end{figure}
\begin{figure}
\includegraphics[width=10cm]{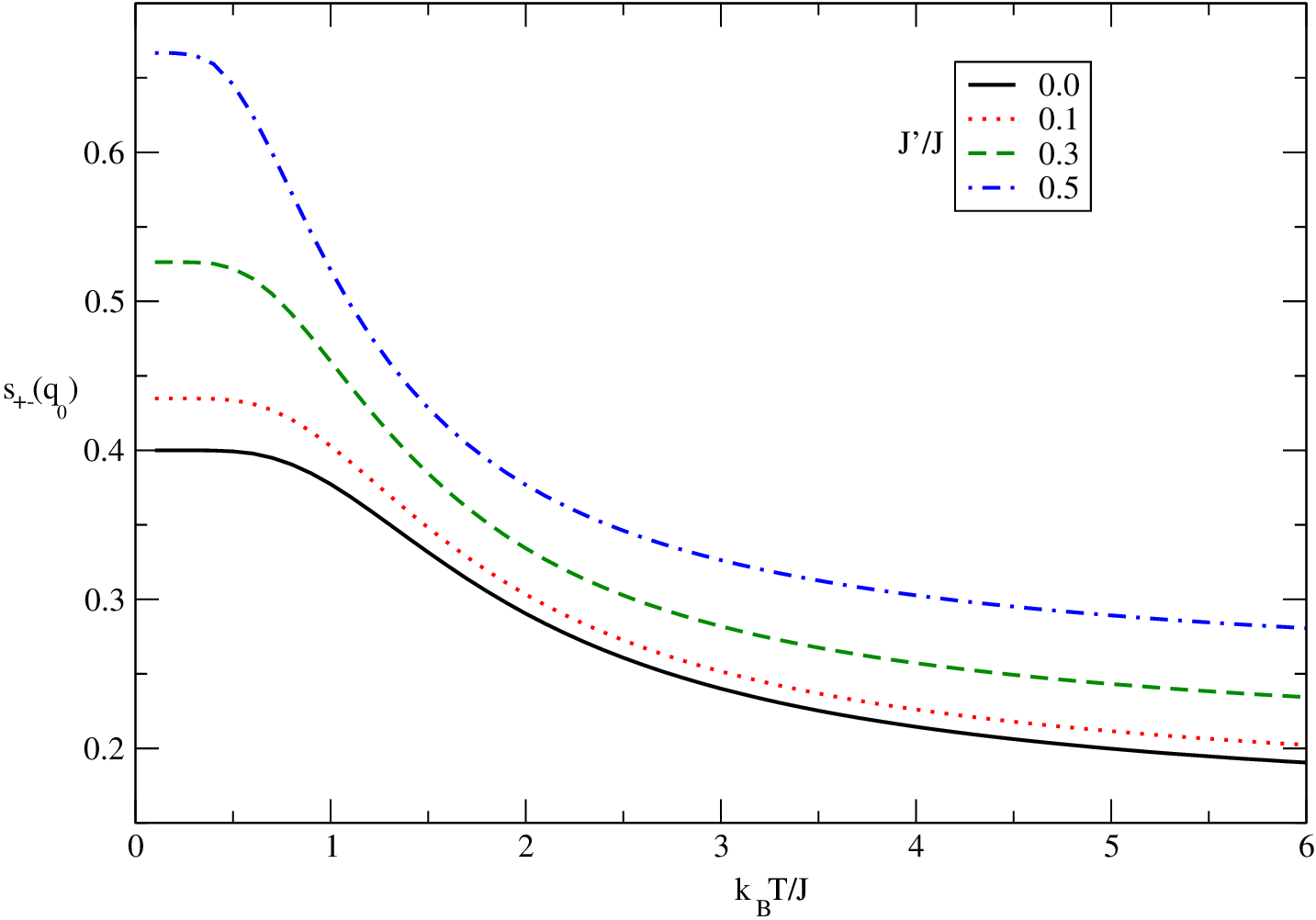}
\caption{\label{fig9}Transverse static spin structure factor $s_{+-}({\bf q}_{0},T)$ as a function normalized temperature $k_{B}T/J$ 
for different values of next nearest neighbor exchange coupling constant $J'/J$ for $D/J=0.2$. Also normalized magnetic field is set to 
$g\mu_{B}B/J=6.0$.}
\end{figure}
\begin{figure}
\includegraphics[width=10cm]{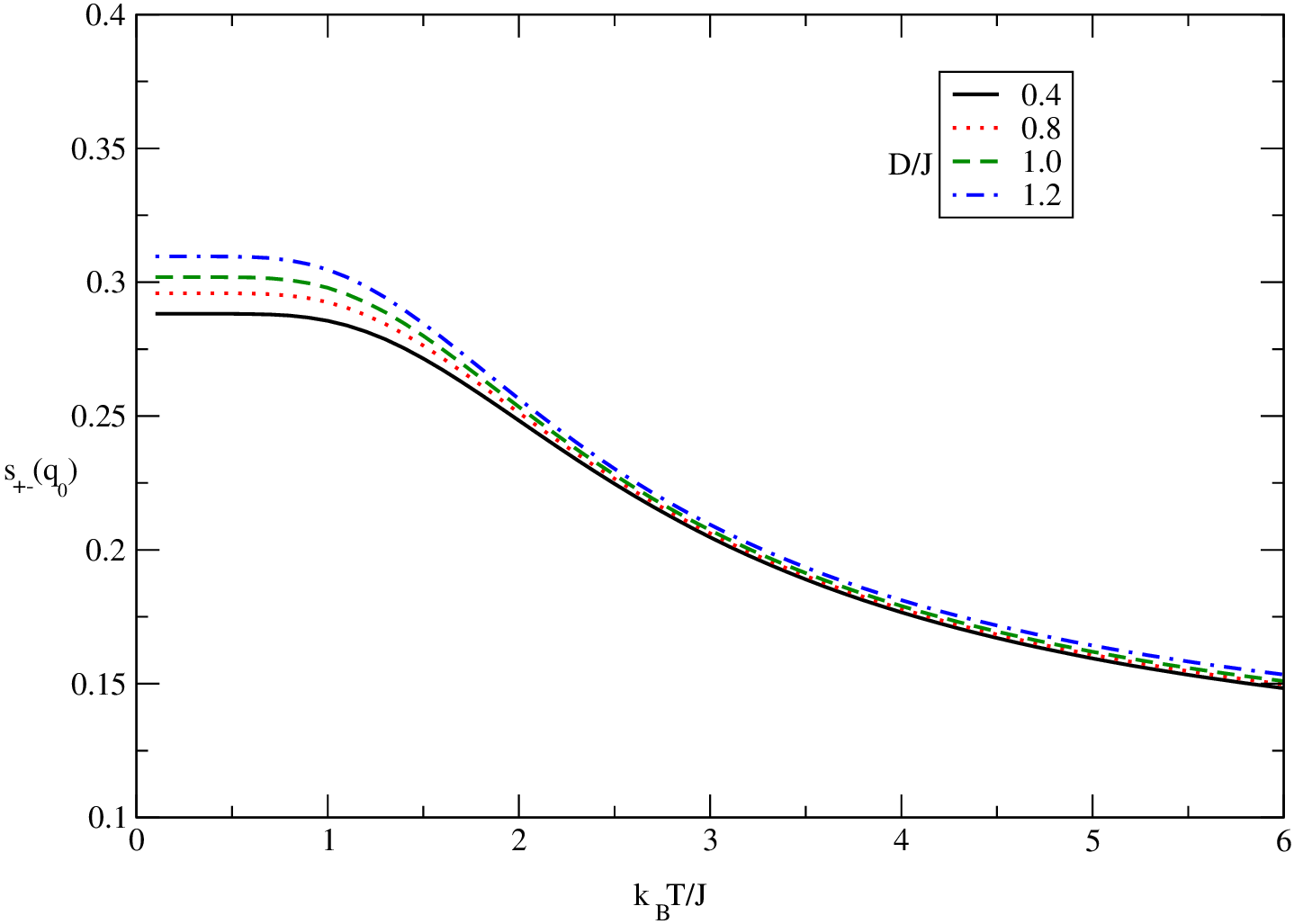}
\caption{\label{fig10}Transverse static spin structure factor $s_{+-}({\bf q}_{0},T)$ as a function normalized temperature $k_{B}T/J$ 
for different values of Dzyaloshinskii-Moriya interaction strength $D/J$ for $J'/J=0.2$. Also normalized magnetic field is set to 
$g\mu_{B}B/J=6.0$.}
\end{figure}

\end{document}